\documentclass[twocolumn,showpacs,preprintnumbers]{revtex4}
\usepackage{amssymb}
\usepackage{amsmath}
\usepackage{graphicx}
\usepackage{dcolumn}
\usepackage{bm}

\begin{document}

\title{Remote Preparation of a Qudit Using Maximally Entangled States of
Qubits}
\author{Chang-shui Yu$^{1}$}
\author{He-shan Song$^{1}$}
\email{hssong@dlut.edu.cn}
\author{Ya-hong Wang$^{1,2}$}
\affiliation{$^{1}$Department of Physics, Dalian University of Technology,\\
Dalian 116024, P. R. China}
\affiliation{$^{2}$School of information science and engineering, Dalian Institute of
Light Industry,\\
Dalian 116034, P. R. China}
\date{\today }

\begin{abstract}
Known quantum pure states of a qudit can be remotely prepared onto a group
of particles of qubits exactly or probabilistically with the aid of
two-level Einstein-Podolsky-Rosen states. We present a protocol for such
kind of remote state preparation. We are mainly focused on the remote
preparation of the ensembles of equatorial states and those of states in
real Hilbert space. In particular, a kind of states of qudits in real
Hilbert space have been shown to be remotely prepared in faith without the
limitation of the input space dimension.
\end{abstract}

\maketitle

\begin{center}
\textbf{I. INTRODUCTION}
\end{center}

Quantum teleportation, as a surprising discovery in quantum information
theory by Bennett et al [1], is the teleportation of an unknown quantum
state from one place to another without physically sending the particle.
E.g. If Alice and Bob share an Einstein-Podolsky-Rosen (EPR) pair, Alice can
teleport a qubit to Bob by first carrying out a Bell-state measurement on
the qubit and one particle of the EPR pair, then sending two bits of
classical information to Bob, who in turn can perform a corresponding
unitary operation on his particle (the other particle of the EPR pair) to
get the state Alice wants to teleport. As a similar case, remote state
preparation (RSP) can be understood as " the teleportation of a known
quantum state". However, the difference between them are in that, 1) in RSP
Alice knows the state that she wants Bob to prepare, in particular, Alice
need not own the state, but only know the information about the state, while
in teleportation Alice must own the teleported state, but she need not know
the state; 2) in RSP, the required resource can be traded off between
classical communication cost and entanglement cost while in quantum
teleportation, two bits of forward classical communication and one ebit of
entanglement (an EPR pair) per teleported qubit are both necessary and
sufficient, and neither resource can be traded off against the other [2,3].

RSP has attracted many attentions in recent years [3-11]. Bennett et al have
generalized RSP for arbitrary qubits, higher dimensional Hilbert spaces and
also of entangled systems [3]. The exact and minimal resource consuming RSP
protocol is generalized to higher dimension by Zeng and Zhang [4]. Ming-Yong
Ye et al consider the faithful remote state preparation using finite
classical bits and a non-maximally entangled state [3]. P. Agrawal et al [5]
discuss the exact remote state preparation protocol of special ensembles of
qubits at multiple locations and also present generalization of their
protocol for higher dimensional Hilbert space systems for multi-parties.
Berry and Sanders [6] has studied remote preparation of ensemble of mixed
states. Simultaneously, RSP for qubits [7] have been implemented in
experiment [8,9]. For RSP protocols in higher dimensional Hilbert space, one
should note that the analogue of singlet EPR pair (multi-level EPR pair)
including the counterpart in Ref. [3] is necessary. However, the distinct
disadvantage is that multi-level EPR pairs is very difficult to prepare. On
the contrary, the preparation of the EPR pairs can be realized by different
schemes [12,13]. A more natural question is that, since a state of qudit can
be teleported on a group of particles of qubits [14], whether the similar
version is suitable for RSP.

In this paper, we present protocols to remotely prepare some qudits mainly
including the ensembles of the equatorial states and those of the states in
real Hilbert space exactly or probabilistically, using EPR pairs. The key to
successfully implementing RSP is that the two states in different Hilbert
spaces (a lower dimensional space and a higher dimensional one) can be
considered to be equivalent provided that the coefficients of one state
expanded in the lower dimensional space are the same to those of the other
state expanded in some a subspace (The dimension is the same to the former.)
of the higher dimensional space. We find that the ensembles of states in
real Hilbert space can be remotely prepared exactly, while only ensembles of
the equatorial states with equal input and output space dimensions can be
remotely prepared exactly. It is interesting that a kind of states of qudits
in real Hilbert space can be remotely prepared without the limitation of the
input space dimension, which is different from that in Ref. [4]. The paper
is organized as follows. In section II, we study the RSP for the ensembles
of the equatorial states; In section III, we study the RSP of the ensembles
of states in real Hilbert space. The conclusion is drawn in Section IV.

\begin{center}
\textbf{II. RSP FOR THE ENSEMBLES OF EQUATORIAL STATES}
\end{center}

At first, it is necessary to rigorously describe the equivalence of the
states in different Hilbert spaces, as mentioned in Introduction and implied
in Ref. [14]. In other words, how do we understand a qudit has been prepared
onto a group of quantum states of qubits? Consider two Hilbert spaces
denoted by $H_{1}$ and $H_{2}$ with dimensions $D_{1}$ and $D_{2}$, $%
D_{1}<D_{2}$. A state $\phi _{1}$ defined in $H_{1}$ can be expanded based
on $D_{1}$ orthogonal and complete bases with $\alpha _{j}$, $j=1,2,\cdot
\cdot \cdot D_{1},$ corresponding to the expanded coefficients. Therefore,
for the given bases, $\phi _{1}$ can be denoted by
\begin{equation*}
\phi _{1}\longmapsto (\alpha _{1},\alpha _{2},\cdot \cdot \cdot ,\alpha
_{D_{1}}).
\end{equation*}%
Another state $\phi _{2}$ defined in any $D_{1}$-dimensional subspace of $%
H_{2}$ can also be expanded in the subspace based on $D_{1}$ orthogonal and
complete bases, with $\beta _{j}$, $j=1,2,\cdot \cdot \cdot D_{1}$ denoting
the expanded coefficients. Analogously, $\phi _{2}$ can be given by
\begin{equation*}
\phi _{2}\longmapsto (\beta _{1},\beta _{2},\cdot \cdot \cdot ,\beta
_{D_{1}}).
\end{equation*}%
One can say that $\phi _{1}$ and $\phi _{2}$ are equivalent if and only
\begin{equation*}
(\alpha _{1},\alpha _{2},\cdot \cdot \cdot ,\alpha _{D_{1}})=(\beta
_{1},\beta _{2},\cdot \cdot \cdot ,\beta _{D_{1}}).
\end{equation*}%
It is obvious that, if there exists such an equivalent state defined in the 2%
$^{n}$-dimensional Hilbert space to a qudit defined in $D_{1}$ dimension,
the minimum $n$ is the integer that satisfies $1+\log _{2}D_{1}\geq n\geq
\log _{2}D_{1}.$

Now, we begin our protocol for RSP. Here we suppose that Alice wants to
remotely prepare a known $s$-level equatorial state at Bob's location with
the state written by%
\begin{equation*}
\left\vert \psi \right\rangle =\frac{1}{\sqrt{s}}\sum_{j=0}^{s-1}e^{i\varphi
_{j}}\left\vert j\right\rangle ,
\end{equation*}%
where $\left\vert j\right\rangle $ is the computational basis and $\varphi
_{j}$ is real. For this purpose, she needs to share $L$ EPR pairs with Bob,
where $L$ is limited by
\begin{equation}
1+\log _{2}s\geq L\geq \log _{2}s.
\end{equation}%
In other words, Alice and Bob have constructed $L$ quantum channels every
one of which is an EPR pair. The quantum channel can be written by%
\begin{equation}
\left\vert \Phi \right\rangle =\left[ \frac{1}{\sqrt{2}}\left( \left\vert
00\right\rangle +\left\vert 11\right\rangle \right) \right] ^{\otimes L},
\end{equation}%
where the EPR pair chosen is $\frac{1}{\sqrt{2}}\left( \left\vert
00\right\rangle +\left\vert 11\right\rangle \right) $. Expanding eq. (2), $%
\left\vert \Phi \right\rangle $ can be rewritten by%
\begin{eqnarray}
\left\vert \Phi \right\rangle &=&\frac{1}{2^{L/2}}(\left( \left\vert
0\right\rangle ^{\otimes L}\right) _{A}\otimes \left( \left\vert
0\right\rangle ^{\otimes L}\right) _{B} \\
&&+\left( \left\vert 0\right\rangle ^{\otimes \left( L-1\right) }\left\vert
1\right\rangle \right) _{A}\otimes \left( \left\vert 0\right\rangle
^{\otimes \left( L-1\right) }\left\vert 1\right\rangle \right) _{B}  \notag
\\
&&+\cdot \cdot \cdot +\left( \left\vert 1\right\rangle ^{\otimes L}\right)
_{A}\otimes \left( \left\vert 1\right\rangle ^{\otimes L}\right) _{B}),
\notag
\end{eqnarray}%
where $($ $)_{A}$ denotes the particles at Alice's location and $($ $)_{B}$
corresponds to Bob's. Since any number can be expressed in binary form, a
given binary number can be written as a decimal one. In this sense, the
quantum channel $\left\vert \Phi \right\rangle $ can be considered to be
written in binary \ form. Correspondingly, the channel in decimal form can
be given by
\begin{eqnarray}
\left\vert \Phi \right\rangle &=&\frac{1}{2^{L/2}}(\left( \left\vert
0\right\rangle \right) _{A}\otimes \left( \left\vert 0\right\rangle \right)
_{B}+\left( \left\vert 1\right\rangle \right) _{A}\otimes \left( \left\vert
1\right\rangle \right) _{B} \\
&&+\cdot \cdot \cdot +\left( \left\vert 2^{L}-1\right\rangle \right)
_{A}\otimes \left( \left\vert 2^{L}-1\right\rangle \right) _{B}).  \notag
\end{eqnarray}%
Therefore, $\left\vert \Phi \right\rangle $ can be considered as a $2^{L}$%
-level EPR pair. In principle, Alice is able to perform any unitary
transformation and measurement on the composite system of the $L$ particles
at her location. Due to $s\leq 2^{L}$, the equatorial state to be prepared
can be expanded in any subsystem of the $2^{L}$ dimensional space. Without
loss of the generality, one can choose the subsystem expanded by the basis
vectors $\{\left\vert 0\right\rangle _{A}$, $\left\vert 1\right\rangle _{A}$%
, $\cdot \cdot \cdot $, $\left\vert s-1\right\rangle _{A}\}$. In the
subsystem, Alice applies a unitary transformation $U^{\prime }$ on the basis
vectors of the subsystem then she obtains%
\begin{equation}
U^{\prime }\left( \left\vert k\right\rangle \right) _{A}=\left\vert \psi
_{k}\right\rangle =\frac{1}{\sqrt{s}}\sum_{j=0}^{s-1}e^{i(2kj\pi
)/s}e^{i\varphi _{j}}\left\vert j\right\rangle ,
\end{equation}%
where $\varphi _{0}=0$. Namely, Alice performs a unitary transformation $U$
on her $L$ particles where $U=U^{\prime }\oplus I_{\alpha }$ with $I_{\alpha
}$ the identity of $\alpha \left( =2^{L}-s\right) $ dimension. After the
transformation, the quantum channel is expressed in form by%
\begin{equation}
\left\vert \Phi \right\rangle =\frac{1}{2^{L/2}}\left(
\sum_{k=0}^{s-1}e^{i\chi _{k}}\left\vert k\right\rangle _{A}\otimes
\left\vert \psi _{k}\right\rangle _{B}+\sum_{k=s}^{2^{L}-1}\left\vert
k\right\rangle _{A}\left\vert k\right\rangle _{B}\right) ,
\end{equation}%
where $\chi _{k}$ are the corresponding phase factors, and not given
explicitly. Note that $\left\vert k\right\rangle $ are written in decimal
form. One can change them back to binary form. Now, Alice performs
single-particle measurements on her particles in the basis $\{\left\vert
0\right\rangle $, $\left\vert 1\right\rangle \}$. It is obvious that if
Alice obtains the outcomes $\left\vert k\right\rangle _{A}$ (decimal form), $%
k\leq s-1$, she has to convey to Bob by classical communication whether to
apply the corresponding unitary transformation
\begin{equation}
V_{k}=diag(1,e^{i(2k\pi )/s},e^{i(4k\pi )/s}),\cdot \cdot \cdot ,e^{i(2k\pi
)(s-1)/s})\oplus I_{\alpha }
\end{equation}%
on his $L$ particles or do nothing. Based on the analysis at the beginning,
Alice can believe that the $s$-level quantum state has been remotely
prepared on the $L$ two-level particles at Bob's location. However, if Alice
obtains the outcomes $\left\vert k\right\rangle _{A}$, $k>s-1$, she will
have to inform Bob of the failing RSP. The successful probability is $%
s/2^{L}\times 100\%>50\%$, due to the limitation (1). The number of cbits
used should be log$_{2}(s+1)$ if $s\neq 2^{L}$, otherwise log$_{2}s$,
because it is not necessary for Bob to know which measurement outcome leads
to the failure. It is obvious that the state is prepared exactly by the
protocol only if $s=2^{L}.$Compared with teleportation, the $s$-level
quantum state can be exactly teleported by $L$ EPR pairs all the time, but
the number of cbits used has to be log$_{2}s+L$.

\begin{center}
{\textbf{III. RSP FOR THE ENSEMBLES OF STATES IN REAL HILBERT SPACE}}
\end{center}

At first, let us recall the RSP protocol proposed in Ref. [4] given in our
version, where Alice wants to remotely prepare a qudit defined in real $s$%
-dimensional Hilbert space which is expressed by%
\begin{equation}
\left\vert \psi \right\rangle =\sum_{j=0}^{s-1}\alpha _{j}\left\vert
j\right\rangle ,
\end{equation}%
with the matrix notation given by $\Psi =[\alpha _{0},\alpha _{1},\cdot
\cdot \cdot ,\alpha _{s-1}]^{T}$, where $\alpha _{j}$ is real, $%
\sum_{j=0}^{s-1}\alpha _{j}^{2}=1$ and superscript $T$ denotes transpose.
The quantum channel, i.e. the maximally entangled state shared by Alice and
Bob is
\begin{equation*}
\left\vert \Phi \right\rangle _{AB}=\frac{1}{\sqrt{s}}\sum_{j=0}^{s-1}\left%
\vert j\right\rangle _{A}\otimes \left\vert j\right\rangle _{B}.
\end{equation*}%
For a feasible minimum RSP task, there should exist $s$ unitary operators $%
V_{j},j=0,1,\cdot \cdot \cdot ,s-1$, independent of $\Psi $ such that all
the vectors given by $V_{j}\Psi $ are orthogonal each other. To realize the
RSP, Alice first performs a unitary operation $U$ given by
\begin{equation*}
U=[V_{0}\Psi ,V_{1}\Psi ,\cdot \cdot \cdot ,V_{s-1}\Psi ]^{T},
\end{equation*}%
on the particle of the channel at her location, then she performs
single-particle measurement with respect to the basis $\{\left\vert
j\right\rangle _{A}\}$. If she obtains the outcome $\left\vert
j\right\rangle _{A}$, she will inform Bob by classical communication. Bob
can perform a corresponding unitary operation $V_{j}^{\dagger }$ on his
particle based on the outcome, then he obtains the state $\left\vert \psi
\right\rangle $. The RSP is completed.

However, not all the qudits can be remotely prepared according to the
minimum RSP. In Ref. [4], the authors have shown that the qudit can be
remotely prepared onto another one at Bob's location only if $s=1,2,4,8$. It
is worthy of being noted that $s=3$ and $s=5,6,7$ are just the special cases
of $s=4$ and $s=8$, respectively, which implies that the quantum channels of
RSP for $s=3$ and $s=5,6,7$ are 4-level maximally entangled state and
8-level one, respectively. The numbers of cbits used for $s=3$ and $s=5,6,7$
are 2 and 3, respectively, which are greater than log$_{2}s$. Hence, they
are not included in the so called minimum RSP.

Next, we firstly generalize the protocol mentioned above to the case with $L$
EPR pairs as quantum channels. Alice and Bob have to share $L$ EPR pairs
with $1+\log _{2}s\geq L\geq \log _{2}s$. The channel is chosen the same to
eq. (2). Consider the decimal form, also eq. (4) can be obtained. Eq. (4) is
the same to a 2$^{L}$-level maximally entangled state in form. When $s\leq 4$%
, the state to be remotely prepared can be written as%
\begin{equation*}
\left\vert \psi \right\rangle =\sum_{j=0}^{3}\alpha _{j}\left\vert
j\right\rangle ,
\end{equation*}%
where $\alpha _{j}=0$ for $j>s-1$. The quantum channel of the two EPR pairs
can be given by
\begin{equation*}
\left\vert \Phi \right\rangle _{AB}=\frac{1}{2}\sum_{j=0}^{3}\left\vert
j\right\rangle _{A}\otimes \left\vert j\right\rangle _{B},
\end{equation*}%
where $j$ is the decimal form of $\{00,01,10,11\}$. Alice operates $%
U=[V_{0}\Psi ,V_{1}\Psi ,\cdot \cdot \cdot ,V_{3}\Psi ]^{T}$ on her two
particles of the channel, where $\Psi =[\alpha _{0},\alpha _{1},\alpha
_{2},\alpha _{3}]^{T}$, $V_{0}=I$, $V_{1}=\left(
\begin{array}{cc}
-i\sigma _{y} & 0 \\
0 & -i\sigma _{y}%
\end{array}%
\right) $, $V_{2}=\left(
\begin{array}{cc}
0 & -\sigma _{z} \\
\sigma _{z} & 0%
\end{array}%
\right) $ and $V_{3}=\left(
\begin{array}{cc}
0 & -\sigma _{x} \\
\sigma _{x} & 0%
\end{array}%
\right) $, with $I$ the identity, $\sigma _{i}$ Pauli matrices. After
Alice's operation, $\left\vert \Phi \right\rangle _{AB}$ is converted into
\begin{equation*}
\left\vert \Phi \right\rangle _{AB}\rightarrow U\left\vert \Phi
\right\rangle _{AB}=\frac{1}{2}\sum_{j=0}^{3}\left\vert j\right\rangle
_{A}\otimes V_{j}\left\vert \psi \right\rangle _{B}.
\end{equation*}%
Alice performs single-particle measurements on her two particles. If Alice
obtains $\left\vert 00\right\rangle $(changed back to binary form), Bob does
nothing. If Alice obtains $\left\vert 01\right\rangle $, Bob operates $%
V_{1}^{\dagger }$ on his two particles. The others are analogous. What's
more, one will find that $V_{j}^{\dagger }$ can be written in the form of $%
M_{j}\otimes N_{j}$. Therefore, after Bob receives Alice's outcome, he can
operate $M_{j}$ and $N_{j}$ on his two particles, respectively. When $%
4<s\leq 8$, Alice and Bob have to share three EPR pairs as quantum channel.
According to the analogous procedure described above, the corresponding
qudit to be prepared can be remotely prepared exactly onto three particles
at Bob's location. The unitary operation $V_{j}$ needed are the same to
those for 8-dimensional RSP in Ref. [4]. However, unlike the case in four
dimension, not all $V_{j}$, $j=1,2,\cdot \cdot \cdot ,8$, can be decomposed
into direct product forms, hence Bob has to perform collective unitary
operations on his three particles based on Alice's outcomes. The number of
cbits used is $L$ for $s\leq 8$, while that for teleportation is log$_{2}s+L$%
. The difference between the present protocol and that in Ref. [4] are
mainly in that:\ In Ref. [4], unitary operations on single particle are only
needed, but in the present protocol, collective unitary operations on
multiple particles have to be preformed.

It is very interesting that a kind of states in real Hilbert space can also
be remotely prepared exactly onto a group of two-level particles without the
limitation of the input space dimension.

\textbf{Case 1.} Consider the $s$-level state to be prepared given by eq.
(8), $s>8$, which can always be expressed in binary form by%
\begin{equation}
\left\vert \psi \right\rangle =\left\vert \psi \right\rangle _{1,2,\cdot
\cdot \cdot ,L}=\sum_{i,j,\cdot \cdot \cdot ,k=0}^{1}\alpha _{ij\cdot \cdot
\cdot k}\underset{L}{\underbrace{\left\vert i\right\rangle \otimes
\left\vert j\right\rangle \otimes \cdot \cdot \cdot \otimes \left\vert
k\right\rangle }},
\end{equation}%
where $L$ denotes the number of EPR pairs as the quantum channel and $1+\log
_{2}s\geq L\geq \log _{2}s$. Furthermore, $\alpha _{ij\cdot \cdot \cdot k}=0$
if $i\times 2^{0}+j\times 2^{1}+\cdot \cdot \cdot +k\times 2^{L-1}>s-1$.
From the mathematical form of eq. (9), $\left\vert \psi \right\rangle $ can
be considered as a $L$-particle quantum state. In this sense, $\left\vert
\psi \right\rangle $ is denoted by $\left\vert \psi \right\rangle
_{1,2,\cdot \cdot \cdot ,L}$, the density matrix of which can be given by $%
\rho _{1,2,\cdot \cdot \cdot ,L}$. From the viewpoint of a multipartite
quantum state, one can give the reduced density matrix by tracing over one
or several of the $L$ subsystems. As we know, a multi-particle pure state is
fully separable if and only if all the reduced density matrices tracing over
$L-1$ subsystems are pure. Namely,
\begin{equation}
\sqrt{2\sum_{r=1}^{L}(1-tr\rho _{r}^{2})}=0,
\end{equation}%
where $\rho _{r}$ is the reduced density matrix of the $r$th subsystem
[15,16]. If we assume $\left\vert \psi \right\rangle _{1,2,\cdot \cdot \cdot
,L}$ is fully separable, $\left\vert \psi \right\rangle _{1,2,\cdot \cdot
\cdot ,L}$ can be written by%
\begin{equation}
\left\vert \psi \right\rangle _{1,2,\cdot \cdot \cdot ,L}=\left\vert \psi
\right\rangle _{1}\otimes \left\vert \psi \right\rangle _{2}\otimes \cdot
\cdot \cdot \otimes \left\vert \psi \right\rangle _{L},
\end{equation}%
where $\left\vert \psi \right\rangle _{i}$ are two-level quantum states,
formally given by
\begin{equation*}
\left\vert \psi \right\rangle _{i}=a_{i}\left\vert 0\right\rangle
+b_{i}\left\vert 1\right\rangle ,a_{i}^{2}+b_{i}^{2}=1,
\end{equation*}%
$a_{i}$ and $b_{i}$ are determined by eq. (9) and eq. (11), and $\alpha
_{ij\cdot \cdot \cdot k}$ are limited by eq. (10).

Since Alice has known all the information on $\left\vert \psi \right\rangle $%
, to remotely prepare $\left\vert \psi \right\rangle $ at Bob's location,
Alice can select the $i$th EPR pair shared with Bob and carry out
measurement on her particle by projecting onto the basis given by%
\begin{eqnarray*}
\left\vert \psi \right\rangle _{i} &=&a_{i}\left\vert 0\right\rangle
+b_{i}\left\vert 1\right\rangle , \\
\left\vert \bar{\psi}\right\rangle _{i} &=&b_{i}\left\vert 0\right\rangle
-a_{i}\left\vert 1\right\rangle .
\end{eqnarray*}%
Then Alice informs Bob of her outcomes, and Bob has to perform a
corresponding unitary transformation $U_{i}=\left(
\begin{array}{cc}
0 & -1 \\
1 & 0%
\end{array}%
\right) $ on his particle of the $i$th EPR pair, if he receives the message $%
\left\vert \bar{\psi}\right\rangle _{i}$, otherwise do nothing. If the same
job is done for all the $L$ EPR pairs, Bob will obtain a genuine two-level $%
L $-particle state $\left\vert \psi \right\rangle _{1,2,\cdot \cdot \cdot
,L} $ (fully separable), which means that the original $s$-level state of
qudit has been exactly remotely prepared onto $L$ two-level particles. The
number of cbits used is $L$. It is very obvious that the key in this case is
that the RSP for a qudit has been converted into $L\ $RSPs for qubit,
meanwhile RSP for the states of qubit in real Hilbert space can be exactly
implemented all the time [7].

\textbf{Case 2.} Following eq. (9), since $\left\vert \psi \right\rangle
_{1,2,\cdot \cdot \cdot ,L}$ has been considered as an $L$-particle quantum
state, we can group the $L$ subsystems into $\left[ \frac{L}{2}\right] $
parties, with every two subsystems regarded as a party, where $\left[ \frac{L%
}{2}\right] =\left\{
\begin{array}{cc}
\frac{L}{2}, & \text{even }L \\
\frac{L+1}{2}, & \text{odd }L%
\end{array}%
\right. $. Without loss the generality, we rewrite $\left\vert \psi
\right\rangle _{1,2,\cdot \cdot \cdot ,L}$ as%
\begin{eqnarray}
&&\left\vert \psi \right\rangle _{1,2,\cdot \cdot \cdot ,\left[ \frac{L}{2}%
\right] } \\
&=&\sum_{i,j,l,m,\cdot \cdot \cdot ,k=0}^{1}\alpha _{ij\cdot \cdot \cdot k}%
\underset{L}{\underbrace{\left( \left\vert i\right\rangle \otimes \left\vert
j\right\rangle \right) \otimes \left( \left\vert l\right\rangle \otimes
\left\vert m\right\rangle \right) \otimes \cdot \cdot \cdot \otimes
\left\vert k\right\rangle }},  \notag
\end{eqnarray}%
where $(\cdot )$ denotes a party. If $L$ is an even, the subsystems can be
grouped exactly, otherwise there has to be a single subsystem left as a
special party at the end. Thus $\left\vert \psi \right\rangle $ can also be
regarded as a $\left[ \frac{L}{2}\right] $-particle quantum state defined in
$\underset{\left[ \frac{L}{2}\right] }{\underbrace{4\times \cdot \cdot \cdot
\times 4}}$ dimension for even $L$ and in $\underset{\left[ \frac{L}{2}%
\right] }{\underbrace{4\times 4\times \cdot \cdot \cdot \times 4\times 2}}$
dimension for odd $L$. Therefore, we can draw a conclusion as follows. The
original quantum state of qudit can be exactly remotely prepared at Bob's
location, if $\left\vert \psi \right\rangle _{1,2,\cdot \cdot \cdot ,\left[
\frac{L}{2}\right] }$ is fully separable.

The reason for it is simple. If $\left\vert \psi \right\rangle _{1,2,\cdot
\cdot \cdot ,\left[ \frac{L}{2}\right] }$ is fully separable, $\left\vert
\psi \right\rangle _{1,2,\cdot \cdot \cdot ,\left[ \frac{L}{2}\right] }$ can
be rewritten as%
\begin{equation}
\left\vert \psi \right\rangle _{1,2,\cdot \cdot \cdot ,\left[ \frac{L}{2}%
\right] }=\left\vert \varphi \right\rangle _{1}\otimes \left\vert \varphi
\right\rangle _{2}\otimes \cdot \cdot \cdot \otimes \left\vert \varphi
\right\rangle _{\left[ \frac{L}{2}\right] },
\end{equation}%
where $\left\vert \varphi \right\rangle _{i}$ are two-particle quantum
states of qubits (or four-level states) for even $L$, given by%
\begin{equation*}
\left\vert \varphi \right\rangle _{i}=a_{i}\left\vert 00\right\rangle
+b_{i}\left\vert 01\right\rangle +c_{i}\left\vert 10\right\rangle
+d_{i}\left\vert 11\right\rangle ,
\end{equation*}%
with $a_{i}^{2}+b_{i}^{2}+c_{i}^{2}+d_{i}^{2}=1$; For odd $L$, $\left\vert
\varphi \right\rangle _{i},$ $i<\left[ \frac{L}{2}\right] ,$ are defined the
same to those for even $L$, while $\left\vert \varphi \right\rangle _{\left[
\frac{L}{2}\right] }=a_{\left[ \frac{L}{2}\right] }\left\vert 0\right\rangle
+b_{\left[ \frac{L}{2}\right] }\left\vert 1\right\rangle $, $a_{\left[ \frac{%
L}{2}\right] }^{2}+b_{\left[ \frac{L}{2}\right] }^{2}=1$. Alice firstly
divides the $L$ EPR pairs into $\left[ \frac{L}{2}\right] $ groups
corresponding to $\left\vert \varphi \right\rangle _{i}$. Hence RSP for $%
\left\vert \psi \right\rangle _{1,2,\cdot \cdot \cdot ,\left[ \frac{L}{2}%
\right] }$ is converted into $\left[ \frac{L}{2}\right] $ RSPs for
four-level states (a two-level state may be included). Since RSP for the
four-level states in real Hilbert space can be exactly implemented as
mentioned at the beginning of the section, the original state of qudit can
be prepared at Bob's location exactly. The number of cbits used here is also
$L$.

\textbf{Case 3. }Following eq. (9) again, if we group the $L$ subsystems
into $\left[ \frac{L}{3}\right] $ parties, with every three subsystems
regarded as a party, where $\left[ \frac{L}{3}\right] $ denotes the minimal
integer greater than $\frac{L}{3}$. Hence $\left\vert \psi \right\rangle $
can be considered as a $\left[ \frac{L}{3}\right] $-particle quantum state
denoted by $\left\vert \psi \right\rangle _{1,2,\cdot \cdot \cdot ,\left[
\frac{L}{3}\right] }$. Since RSP for the eight-level states in real Hilbert
space can be exactly implemented [4], based on the same procedure to that in
Case 2, we can also draw the conclusion as follows. The original quantum
state of qudit can be remotely prepared exactly at Bob's location, if $%
\left\vert \psi \right\rangle _{1,2,\cdot \cdot \cdot ,\left[ \frac{L}{3}%
\right] }$ is fully separable. The number of cbits used here is $L$ again.

\textbf{Case 4. }The above three cases imply that the $L$ subsystems of $%
\left\vert \psi \right\rangle _{1,2,\cdot \cdot \cdot ,L}$ is divided into
some parties, where each party includes the subsystems with the same
quantity except the subsystems left in the end. If we divided the $L$
subsystems into parties such that every party includes at most three
subsystems, we can also obtain a new multi-particle state (denoted by $\Psi $%
) with its subsystems defined in different dimensions which do not exceed $8$%
. One can find that RSP for the original state $\left\vert \psi
\right\rangle $ can be successfully implemented, if $\Psi $ regarded as a
multi-particle state is fully separable. The reason is that the qudit
defined in $s$ dimensions with $s\leq 8$ can be remotely prepared exactly.
For a given grouping, the number of cbits used here is $L$. Consider
different groupings, the number of cbits used will increase, because Alice
has to inform Bob of how she has divided the $\left\vert \psi \right\rangle
_{1,2,\cdot \cdot \cdot ,L}$.

In fact, for a given $\left\vert \psi \right\rangle $ to be remotely
prepared, Alice can perform any unitary transformation $U_{s}$ on it in mind
(The practical transformation on $\left\vert \psi \right\rangle $ is not
necessary, because it is not the physical state $\left\vert \psi
\right\rangle $ that participates in RSP, but Alice performs some operations
on her particles of the quantum channel based on the information of $%
\left\vert \psi \right\rangle $, which is the difference between RSP and
teleportation mentioned in Introduction), which changes the initial $%
\left\vert \psi \right\rangle $ into $U_{s}\left\vert \psi \right\rangle $.
Therefore, the previous RSP of $\left\vert \psi \right\rangle $ has to be
changed into the RSP of $U_{s}\left\vert \psi \right\rangle $. However, to
realize such a RSP, Alice must inform Bob of what she has done on $%
\left\vert \psi \right\rangle $, i.e. the information of $U_{s}$ by
classical communication. Alternately, Alice and Bob can promise before RSP
task that only one or several given $U_{s}$ are allowed to perform on $%
\left\vert \psi \right\rangle $ by Alice. From the viewpoint that $%
\left\vert \psi \right\rangle $ can be considered as a multi-particle
quantum state $\left\vert \psi \right\rangle _{1,2,\cdot \cdot \cdot ,L}$,
it is possible that $U_{s}$ is not a local unitary transformation. Hence,
the separability of $\left\vert \psi \right\rangle _{1,2,\cdot \cdot \cdot
,L}$ (including all possible groupings mentioned above) will be changed with
$U_{s}$. That is to say, the four cases are dependent on $U_{s}$. Therefore,
so long as Alice can find a $U_{s}$ so that $U_{s}\left\vert \psi
\right\rangle $ satisfies one of the four cases, she can remotely prepare $\left\vert \psi \right\rangle $ successfully onto a group of two-level
particles at Bob's location. However, as mentioned above, Alice has to
inform Bob by classical communication of the $U_{s}$ and her measurement
outcomes, which means that the number of cbits used will be increased with
the increase of the quantity of $U_{s}$ to be employed. However, it is not
expected that the number of cbits used exceeds that used in teleportation.
Hence, the quantity of $U_{s}$ should be restricted before RSP task. In the
realistic scenarios, before RSP, Alice and Bob have to make some limitation
on the channels and Alice's operations such as arranging and numbering
their quantum channels, fixing the concrete form and quantity of the
transformations $U_{s}$, restricting the form of groupings, and so on, which
can effectively decrease the number of cbit used. At least, Alice can only
be allowed to follow the above four cases, which means that $U_{s}$ is not
allowed and only a kind of grouping can be considered. In a word, Alice and
Bob can increase the limitations to decrease the number of cbits used.

\begin{center}
{\textbf{IV. CONCLUSIONS}}
\end{center}

We have presented protocols to remotely prepare the ensembles of the
"equatorial states" and those of the states in real Hilbert space, using EPR
pairs exactly or probabilistically. We find that the ensembles of the $s$%
-level equatorial states can be remotely prepared exactly by $L$ EPR pairs
if the input and output space dimensions are equal, otherwise, they can be
prepared with the probability greater than $1/2$. The number of cbits used
is log$_{2}(s+1)$ if $s\neq 2^{L}$, otherwise log$_{2}s$. The ensembles of
states in real Hilbert space can be remotely prepared exactly all the time
if the input space dimensions are less than or equal to $8$. The number of
cbits used is $L$. Some states with the input space dimensions greater than $%
8$ can also be shown to be exactly remotely prepared on a group of two-level
particles so long as they satisfy one of the all cases mentioned in Section
III.

\begin{center}
{\textbf{V. ACKNOWLEDGEMENT}}
\end{center}

This work was supported by the National Natural Science Foundation
of China, under Grant Nos. 10575017 and 60472017.

\end{document}